\documentstyle[11pt,colordvi]{article}
\input{psfig.sty}

\def\Journal#1#2#3#4{{#1} {\bf #2}, #3 (#4)}

\def\NIMA{{\em Nucl. Instrum. Methods} A}

\def\PLB{{\em Phys. Lett.}  B}
\def\PRL{\em Phys. Rev. Lett.}
\def\PRD{{\em Phys. Rev.} D}

\def\EPJ{{\em Eur. Phys. J.} C} 

\begin{document}
\begin{center}{\Large\bf  Jet Production at CDF}
\footnote{Presented on behalf of CDF collaboration 
at `` Corfu 2001: Summer School and Workshop on High Energy Physics, 
September 1-14, 2001.''}
\\
\vglue0.2cm
              {\Large  Christina Mesropian\\}
                       {CDF Collaboration}
\vglue0.3cm
            {\sl The Rockefeller University\\}
             {\sl    1230 York Avenue\\}
             {\sl     New York, NY  10021\\}
             {\sl      USA\\}
\end{center}
\begin{abstract}
In this talk I present the results from the measurement of the
inclusive jet cross section and strong coupling constant based on 
 the CDF Run 1B data, and discuss prospects for Run 2.
\end{abstract}

\section{Introduction}
Jet production at hadron colliders
provides an excellent opportunity for testing the theory of strong
interactions,  Quantum Chromodynamics (QCD).
QCD has achieved remarkable success in describing hadron 
interactions at short distances (large momentum transfers),
owing to the property of asymptotic freedom.
Processes with large momentum transfer can then be 
described by an expansion in powers of $\alpha_s(\mu)$.
Higher energy collisions produce jets of higher energy particles, while
the value of $\alpha_s$ decreases, improving the validity of
perturbative expansion.  
Since the first measurements of the inclusive jet cross section by UA2,
many experiments studying jet production published
comparisons between
theoretical predictions  and experimental measurements.
In particular, 
results from the CDF Run 1A data (1992-1993) inclusive differential
jet cross section  measurement,
which showed significant excess of the data over  the theoretical
predictions at high $E_T$,
 generated lot of excitement due to  the possibilities of new
physics, and   motivated the re-evaluation of all components of
theoretical calculations.

\begin{figure}
\centerline{\psfig{figure=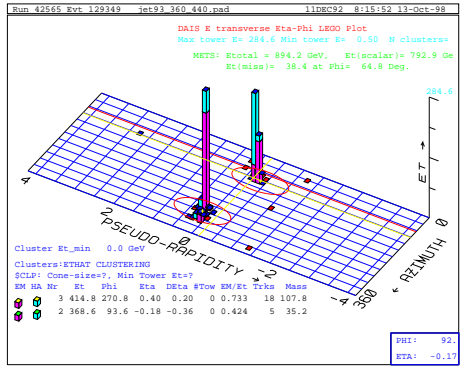,width=6.cm}}
\caption{Lego display of jet event from CDF.}
\label{jets-lego}
\end{figure}

The cornerstone of data to theory comparisons is the question of jet
identification.
In theoretical calculations,  jets are manifestations of 
partons as relatively isolated sprays of energetic hadrons observed in
the final state of high energy collisions.
From the experimental point of view jets are defined as large energy
deposits in a localized group of calorimeter cells, see
Fig.~\ref{jets-lego}.
%-----------
%------------
To minimize the difference between parton level predictions and measured
jet properties, the clustering algorithms are used which could be
implemented for both situations. Jet algorithms start from a list of
``particles'' that are taken to be calorimeter towers in experimental
case and partons for theoretical calculations, and associate clusters
of these particles into jets in such a way that the kinematic properties of
the jets can be related to the properties of the energetic partons
produced in the hard scattering process. 
This way jet algorithms allow
us to see the partons in the final hadronic state. 
Differences in the properties of reconstructed jets when going from
parton to hadronic or calorimeter level are of major concern for a
good jet algorithm.

%-----------
\begin{figure}[h]
\centerline{\hbox{\psfig{figure=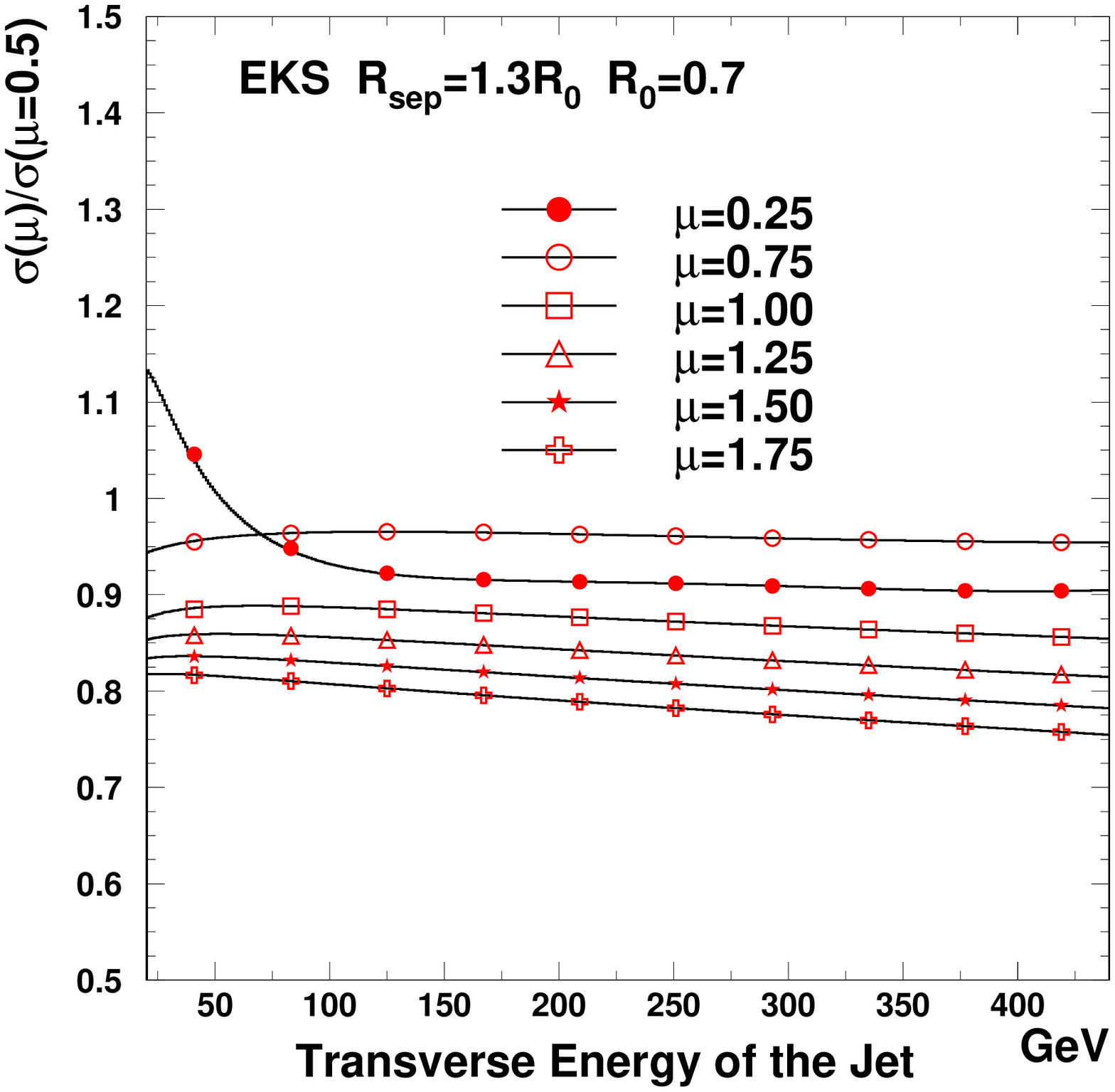,width=7cm}
                  \psfig{figure=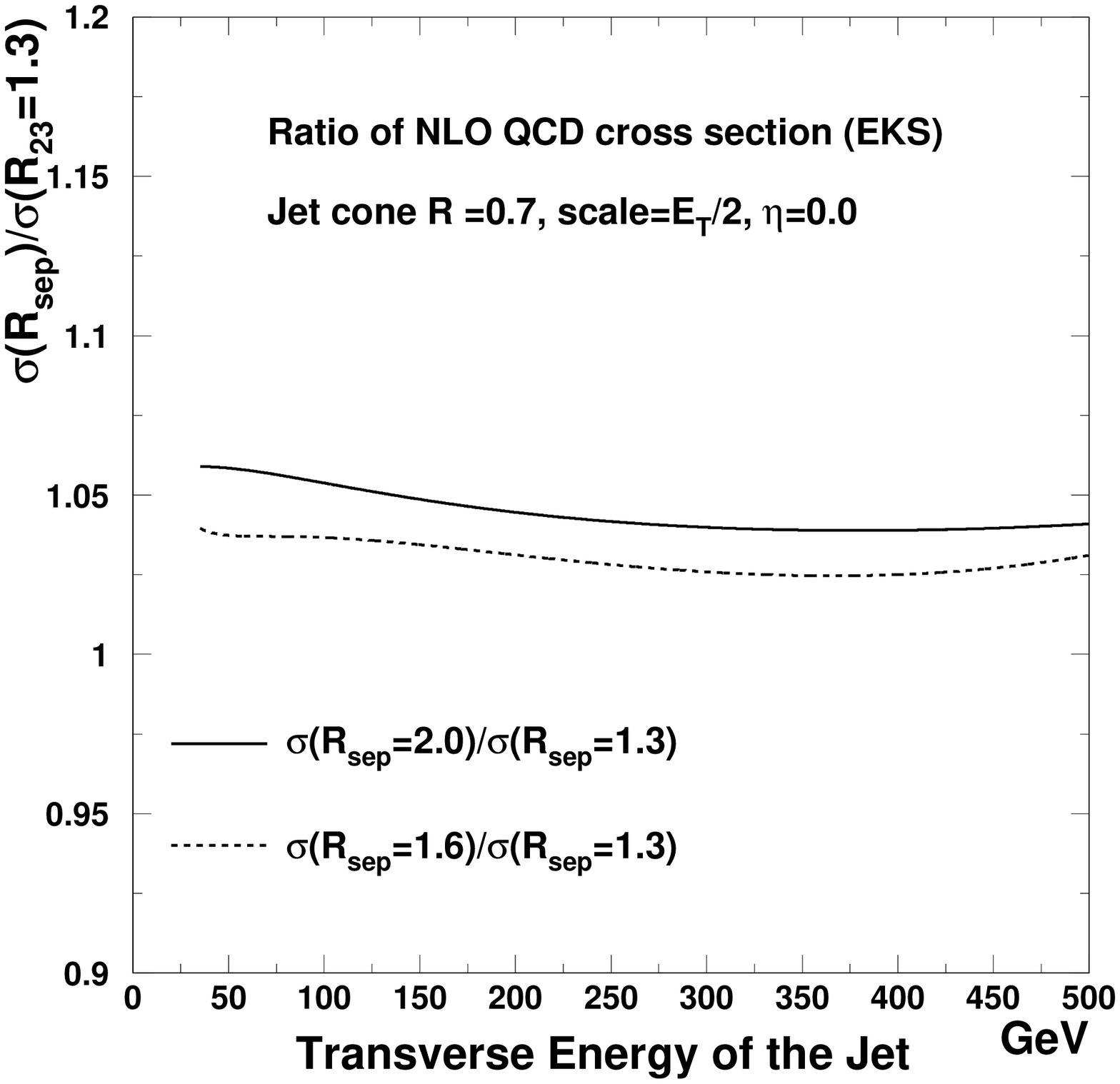,width=7cm}}}
\caption{(left) Variation in theory predictions for different renormalization
scales;
(right)The variation of the inclusive jet cross section for
different ${\cal{R}}_{sep}$ parameters. 
 These calculations used the EKS program.}
\label{inc-jets-mu-renorm}
\end{figure}
%---------------------------------------------------

Historically  cone algorithms were used for
jet measurements in hadron-hadron colliders. 
In Snowmass algorithms,
 two partons are combined into a {\em jet} if
they  fall
within a cone of radius $R$ in $\eta-\phi$ space, where 
$R=\sqrt{(\Delta\eta^2+\Delta\phi^2)}$ and $\Delta\eta$, $\Delta\phi$
are the separation of the partons in pseudo-rapidity and azimuthal
angle. It is further required that the
axis of the cone coincides with the jet direction as defined by $E_T$
weighted centroid of the particles within the cone, 
where $E_T=E\sin\theta$ is the
transverse energy. A similar algorithm with $R=0.7$  is used in the
experimental data analysis by using calorimeter towers instead of
partons. 
The latest  modification of this algorithm when applied to the
theoretical calculations introduces an {\em ad hoc}
parameter ${\cal{R}}_{sep}$~\cite{Ellis-rsep} which is used to regulate the merging and
separation of parton clusters in manner similar to the experimental analysis.
In addition to the difficulty with merging/splitting rules which were
not addressed in the original Snowmass algorithm, cone algorithms
turned out to be sensitive to the soft radiation at the 
NNLO calculations. The second type of algorithm, is so-called $K_T$
algorithm which is infrared and collinear safe to all orders of
calculations and gives close correspondence with jets reconstructed in
calorimeter with jets reconstructed from partons, since it
successively merges
pairs of  ``particles'' in order of increasing relative transverse
momentum. The difficulties of applying this algorithm 
are due to the question of the 
subtraction of the underlying event energy and problem of minimizing
computing time.

There are additional  issues
which affect sensitivity of the
measurements.
First is the limited accuracy of the fixed order perturbative
calculations due in part to the intrinsic uncertainties associated
with the choice of renormalization and factorization scales, $\mu_R$
and $\mu_F$, see Fig.~\ref{inc-jets-mu-renorm}(left).
For the usual range of $\mu=E_T/2$ to $\mu=2E_T$ the variation in
prediction is around 20\%.

The variation in above mentioned 
${\cal{R}}_{sep}$ parameter attributes to some 5-6\%
uncertainty, Fig.~\ref{inc-jets-mu-renorm}(right).
But the 
limited knowledge or parton distribution functions
(PDFs), 
which are obtained from global fits 
to deep inelastic scattering (DIS), Drell-Yan production
and other collider data, 
 are the
largest source of  uncertainty for the calculations.
 The effect of this uncertainty will be discussed later.

\section{Inclusive Jet Cross Section}

The inclusive jet cross section  measurement  is based on a data sample 
of integrated luminosity $87~{\rm pb}^{-1}$ collected
by the Collider Detector at Fermilab (CDF) during the 1994-95 
run  (Run 1b) of the Fermilab Tevatron $\bar pp$ collider operating 
at $\sqrt{s}=1.8$ TeV.
The CDF detector is described elsewhere~\cite{CDF-detector}. 
Details of the measurement of the inclusive jet differential cross 
section can be found in \cite{CDF-Run1b}. 
Briefly, 
jets are reconstructed using the iterative
fixed cone algorithm mentioned above.
%-----------
\begin{figure}[h]
\centerline{\hbox{\psfig{figure=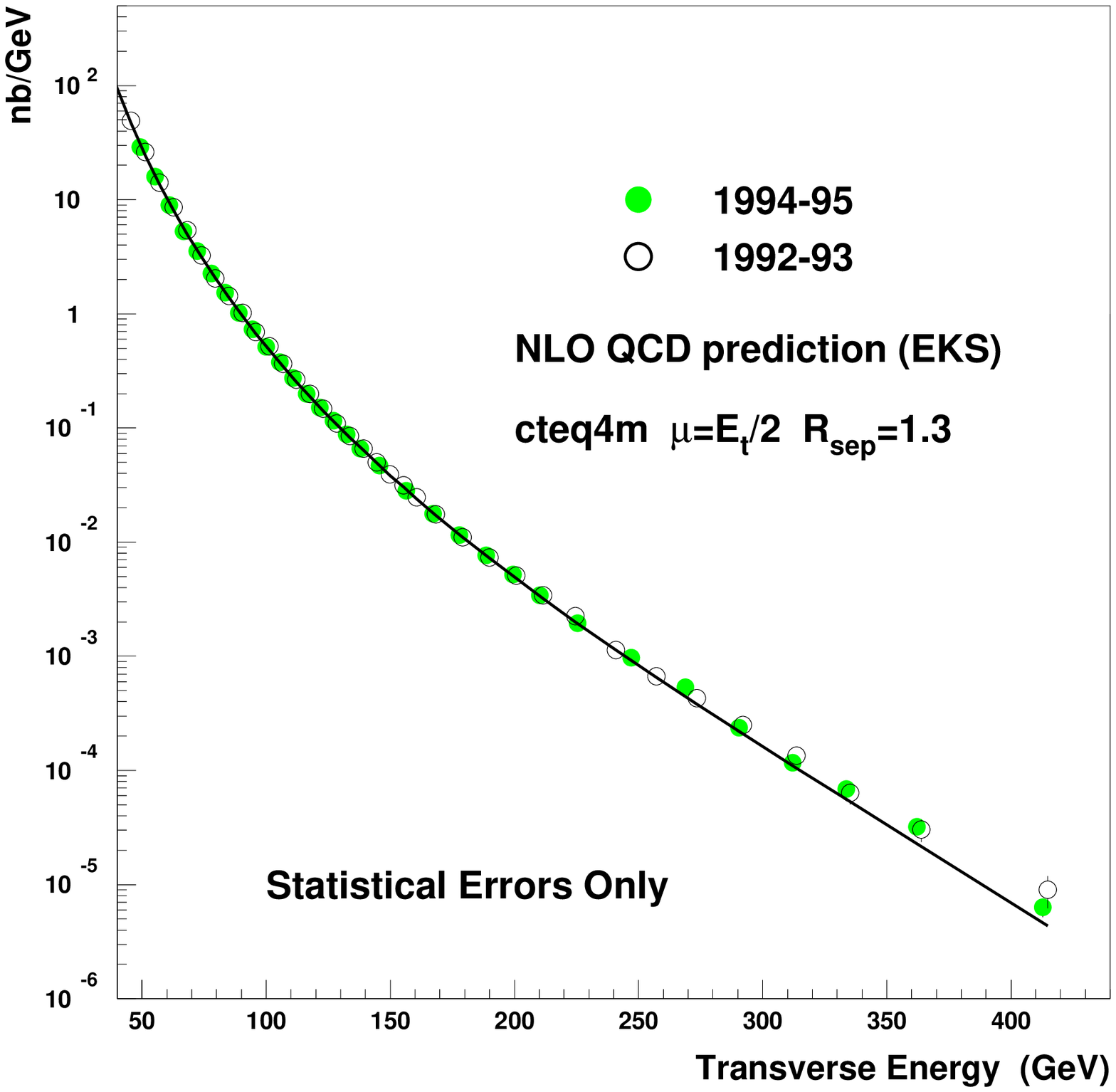,width=7cm}
                  \psfig{figure=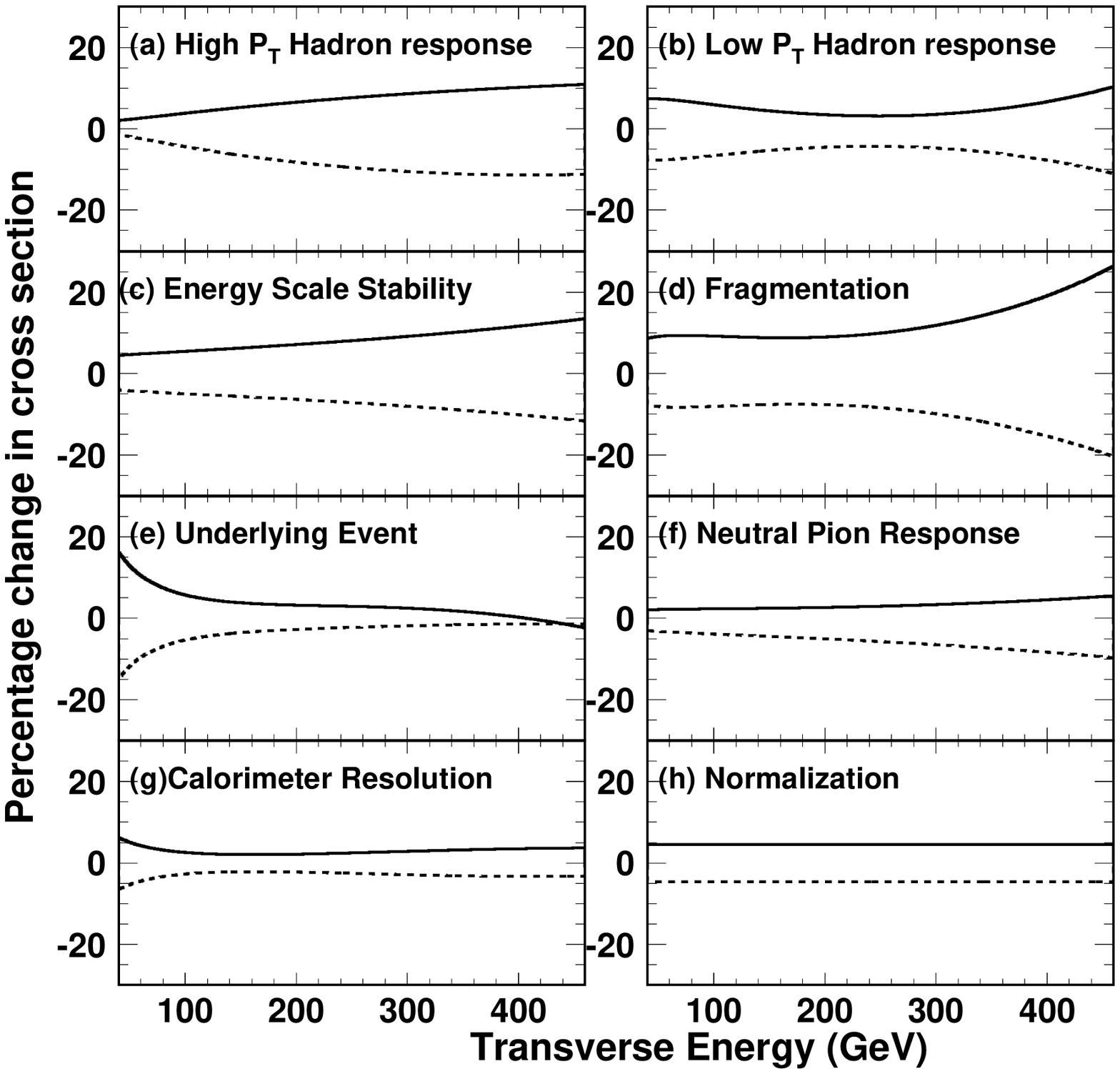,width=7cm}}}
\caption{(left) Inclusive jet cross section from the Run 1B data (1994 to
1995) compared to a QCD prediction and to the published Run 1A data
(1992 to 1993);
(right) The $\pm 1\sigma$ fractional change in cross section due to
the dominate sources of systematic uncertainty.}
\label{inc-jets-1800}
\end{figure}
%---------------------------------------------------
The inclusive jet cross section
includes all jets in an event  in the 
pseudorapidity range $0.1<|\eta|<0.7$.
The measured spectrum is corrected for the calorimeter response, resolution
and the underlying event energy using an iterative unsmearing 
procedure which changes both the energy scale and the normalization
simultaneously.
Figure~\ref{inc-jets-1800}(left) shows the corrected Run 1B cross section
compared to a QCD prediction and Run 1A results.
As we see, the results of Run 1B and Run 1A are in good agreement.

The majority of experimental systematic uncertainty arises from the
uncertainty in the simulation of the response of the detector to jets.
The simulation is tuned to the data for charged hadron response, jet
fragmentation , and $\pi^0$ response.  Additional uncertainty is
associated with the jet energy resolution, the definition of
underlying event, the stability of the detector calibration over the
long running periods and an overall normalization uncertainty from the
luminosity determination. 
For each source 
of systematic uncertainty described above, 
except normalization,
the inclusive jet cross
section was re-evaluated by varying the corresponding parameter
in the detector response by 1 $\sigma$.  For the normalization uncertainty
it was changed by a scale factor, see Fig.~\ref{inc-jets-1800}~(right).

\begin{figure}
\centerline{\psfig{figure=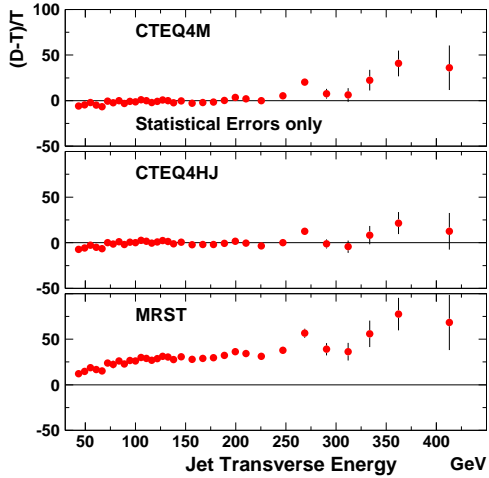,width=7.cm}}
\caption{Run 1B data compared to QCD predictions (EKS, $\mu=E_T/2$,
${\cal{R}}_{sep}=1.3$) using the CTEQ4M, CTEQ4HJ and MRST PDF's. Only
statistical uncertainties are shown on the data points.}
\label{inc-jets-d-t}
\end{figure}

Figure~\ref{inc-jets-d-t} shows the cross section compared
to QCD predictions using three recent PDFs. 
Good agreement is observed using conventional (CTEQ4M and MRST) PDFs
except at high values of transverse energy, starting at 200 GeV, where
an excess is observed. 
Similar deviation have been observed in other CDF jet cross section
measurements such as the dijet mass~\cite{dijet-mass}, 
dijet differential~\cite{dijet}, and $\Sigma E_T$~\cite{sumet}
analyses. 
 One
possible explanation for the excess of high $E_T$ jets is that the
gluon distribution at high $x$ is larger than conventional PDFs have indicated.
 We see that CTEQ4HJ, 
(special PDF set, which was created 
to explore the flexibility in the gluon distribution at high $E_T$, 
and was generated by
including CDF jet data  in the global fit with higher statistical weight
assigned to high $E_T$ points and a new parameterization of 
the gluon distribution~\cite{CTEQ4M}), 
gives the best overall agreement with the data in overall shape and
normalization.
%-----------
%---------------------------------------------------

\section {Strong Coupling Measurement}

The value of $\alpha_s$, a free parameter of QCD, is one of the  
fundamental constants of nature.  Its determination is the essential 
measurement of QCD, and the observation of its evolution, or {\em running}, 
with momentum transfer is one of the key tests of the theory.

The value of $\alpha_s$ is determined by comparing
the jet cross section  with
the next to leading order (NLO)  perturbative QCD 
predictions~\cite{Walter-alphas}.
In the $E_T$ region studied, 
the non-perturbative contributions 
to the inclusive jet cross section are estimated to be 
negligible~\cite{Ellis-rsep}. 
The procedure of extracting $\alpha_s$
can be summarized by the  equation
%--------------------
\begin{equation} 
\label{eq-main}
\frac{d\sigma}{dE_T}=\alpha_s^2(\mu_R)\hat{X}^{(0)}(\mu_F,E_T)\left [1+
\alpha_s(\mu_R)k_1(\mu_R,\mu_F,E_T)\right]
\end{equation}
%---------------------
where 
$\frac{d\sigma}{dE_T}$ is the transverse energy distribution of the inclusive
jets, 
$\mu_R$ and $\mu_F$
are the renormalization and factorization  scales, 
$\alpha_s^2(\mu_R)\hat{X}^{(0)}(\mu_F,E_T)$ is the
 leading order (LO) prediction
for the inclusive jet cross section, and
$\alpha_s^3(\mu_R)\hat{X}^{(0)}(\mu_F,E_T)k_1(\mu_R,\mu_F,E_T)$
the NLO contribution.
Both $\hat{X}^{(0)}$ and $k_1$ are calculated with the 
{\sc jetrad} Monte Carlo program~\cite{JETRAD}.
All calculations are performed in the modified minimal subtraction, 
$\overline{{\rm MS}}$, scheme.
The {\sc jetrad} Monte Carlo
program  has as an input parameters: jet cone radius $R$, ${\cal{R}}_{sep}$,
PDF set and $\mu_R,\mu_F$.

\begin{figure}[h]
\centerline{\hbox{\psfig{figure=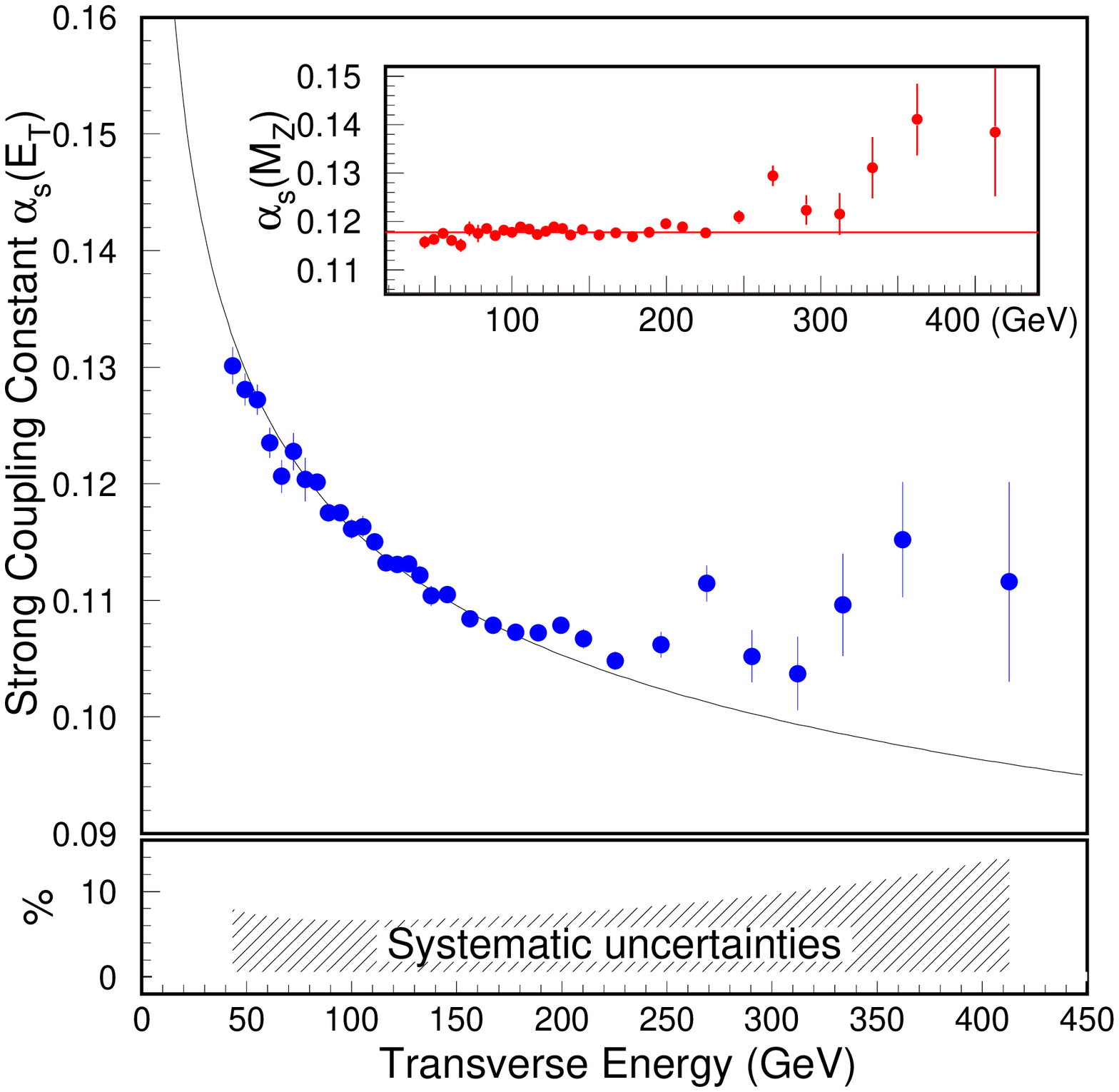,width=7cm}
                  \psfig{figure=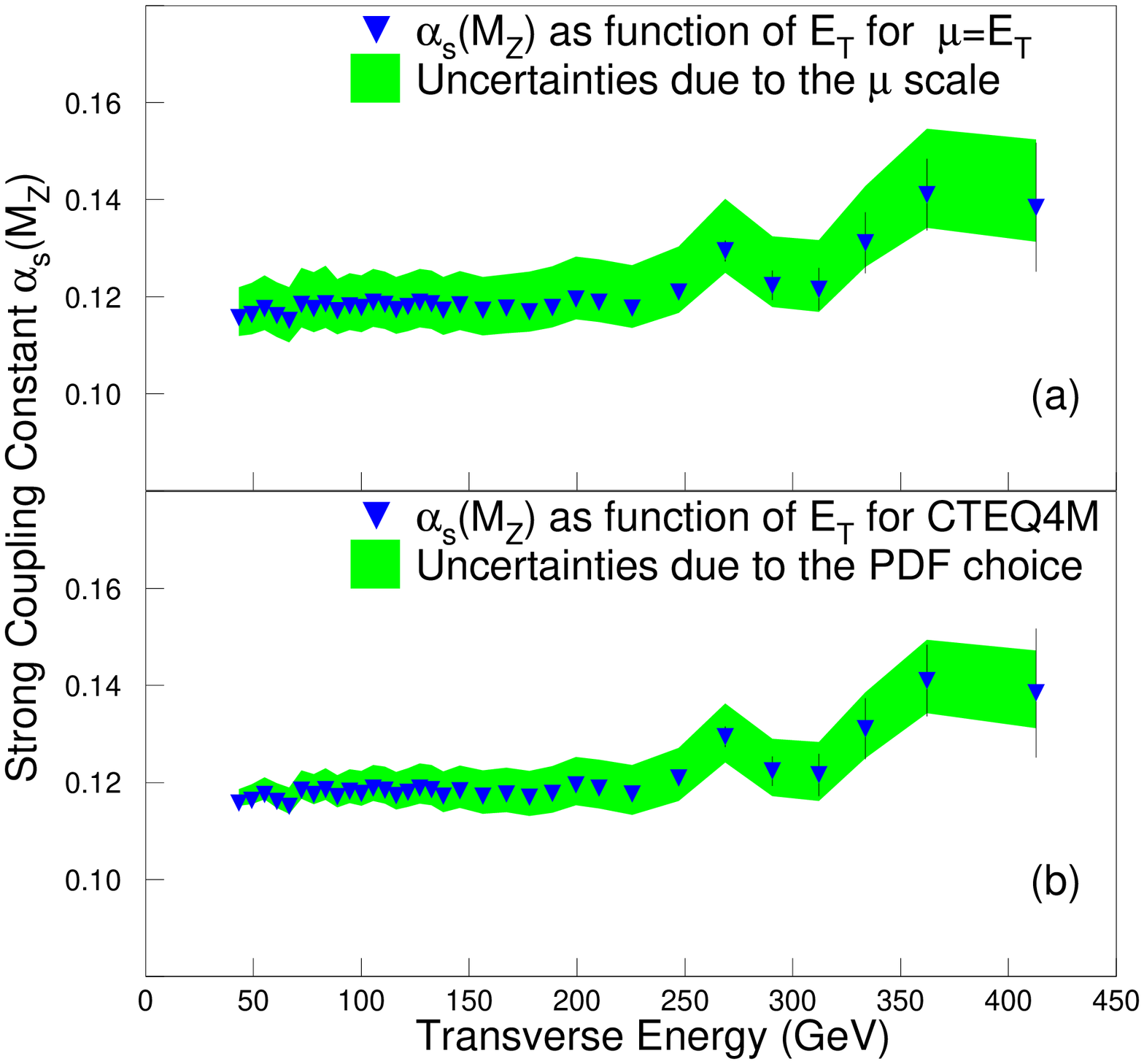,width=7cm}}}
\caption{(left) The strong coupling constant as a function of $E_T$  
for $\mu_R=E_T$ measured using {\sc cteq4m} parton distributions. 
The shaded area shows  the experimental systematic uncertainties. 
The curved line represents the NLO QCD prediction for the evolution of 
$\alpha_s(E_T)$ using $\alpha_s(M_Z)=0.1178$, the average value 
obtained in the region $40<E_T<250$ GeV.
The $\alpha_s(M_Z)$ extracted from $\alpha_s(E_T)$ is shown in the inset
along with the weighted average as the horizontal line;
(right) Uncertainties in $\alpha_s(M_Z)$ due 
to the renormalization scale $\mu$, 
(a), and parton distribution functions, (b).}
\label{alphas_main}
\end{figure}

The inclusive jet data are divided into 33 $E_T$ bins, 
from which we obtain  statistically independent measurements of
$\alpha_s$ for 33 different values of $\mu_R$. 
The $\alpha_s$ values derived for $\mu_R=\mu_F=E_T$ using 
{\sc cteq4m}~\cite{CTEQ4M} parton distribution functions (PDFs) 
are presented in Fig.~\ref{alphas_main} (left). For $E_T$$<$250 GeV, there is 
good agreement with QCD
predictions for the  running of the coupling constant.
The behavior of $\alpha_s$ at higher  $E_T$ values  
is a direct reflection of the 
excess observed  in the inclusive jet cross section~\cite{CDF-Run1b}. 
%---------------------------------------------------
%---------------------------------------------------

The measured values of $\alpha_s(\mu_R)$ are evolved to the  
mass of the $Z^0$ boson, $M_Z$,
by using the solution to the 2-loop renormalization group equation. 
The values of 
$\alpha_s(M_Z)$ for all 33 measurements are shown in the inset of 
Fig.~\ref{alphas_main}(left).
Averaging over the range 40-250 GeV, we obtain~\cite{alphas-prl}
\[\alpha_s(M_Z)=0.1178\pm0.0001{\rm (stat)}.\]
Inclusion of the data with $E_T>$250 GeV results in an increase of 
the average value by  0.0001.

The  experimental systematic uncertainties on the
value of $\alpha_s(M_Z)$ are derived from  
those on the inclusive jet cross section and when 
 summed in quadrature yield   $^{+7\%}_{-8\%}$ uncertainty.

The theoretical uncertainties are mainly due to the
choice of renormalization and factorization scales
and parton distribution functions. The scales $\mu_F$ and $\mu_R$ 
are expected to be of the same order as the characteristic scale of
the process, which in this case is the jet $E_T$. 
We have evaluated the changes in $\alpha_s(M_Z)$ resulting from 
independently varying  $\mu_F$ and  $\mu_R$ from 
$E_T/2$ to $2E_T$ and found that the largest changes occur for $\mu_R=\mu_F$.  
For all results presented in this letter the two scales were set equal.
The sensitivity of the measured $\alpha_s(M_Z)$ to
changes in these scales is indicated by the shaded band in
Fig.~\ref{alphas_main}(right)(a).
Over the $E_T$ range from 40 to 250 GeV, the 
shift in $\alpha_s(M_Z)$ induced by changing the scale 
from $E_T/2$ 
to $2E_T$ 
is approximately $^{+6\%}_{-4\%}$, independently of $E_T$.
We use the {\sc cteq4a} series to study the
$\alpha_s(M_Z)$ dependence on the PDFs. 
In addition, we have studied $\alpha_s(M_Z)$ using 
PDF sets which do not include Tevatron jet results, such as the
{\sc mrst}(g$\uparrow$) set~\cite{MRST}, 
 the {\sc mrsa}$^\prime$ series~\cite{MRSA}, 
and two {\sc mrs-r} sets~\cite{MRSR}. 
The $\chi^2$, calculated by
comparing the data with the theoretical prediction in
the restricted range of 40-250 GeV, is used to quantify 
the agreement.
The minimal  $\chi^2/d.o.f.=1.38$ is obtained for {\sc cteq4m} 
($\alpha_s^{PDF}$=0.116), 
therefore we use this PDF in our final fit.
Excluding the PDFs which have obvious  disagreement
($\chi^2/d.o.f.\ge 5$), we estimate the uncertainty on the
$\alpha_s(M_Z)$ due to the PDF choice to be $\pm 5\%$, 
see Fig.~\ref{alphas_main}(right)(b). 

The resulting  value of $\alpha_s(M_Z)$ 
is in good agreement with the world average 
 $\alpha_s(M_Z)=0.1181\pm 0.0020$.
But what is more important running of $\alpha_s$ is shown over a very
large energy range in the single experiment.

\section{Dijet Triply Differential  Cross Section}

\begin{figure}[h]
\centerline{\hbox{\psfig{figure=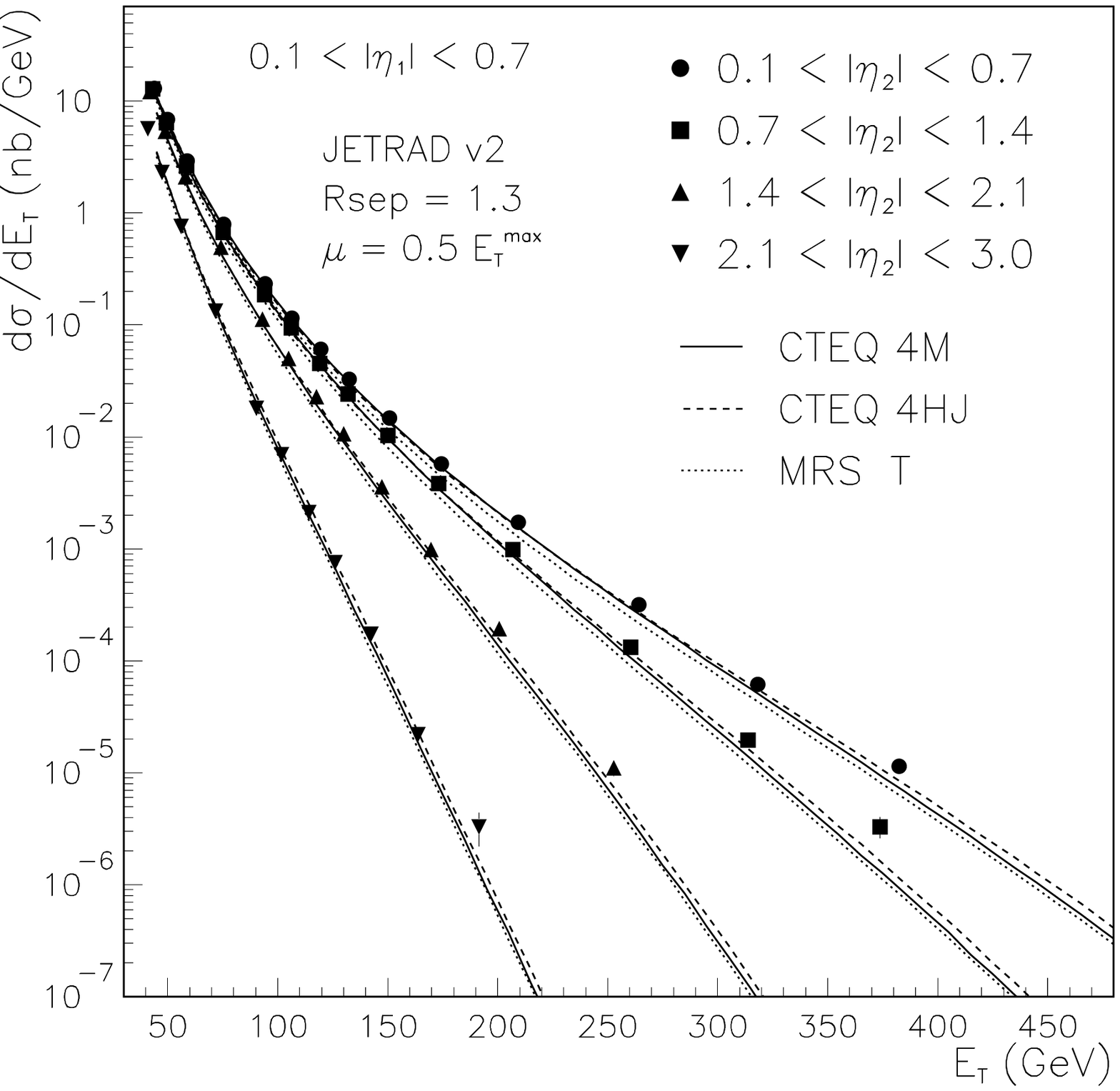,width=7cm}
                  \psfig{figure=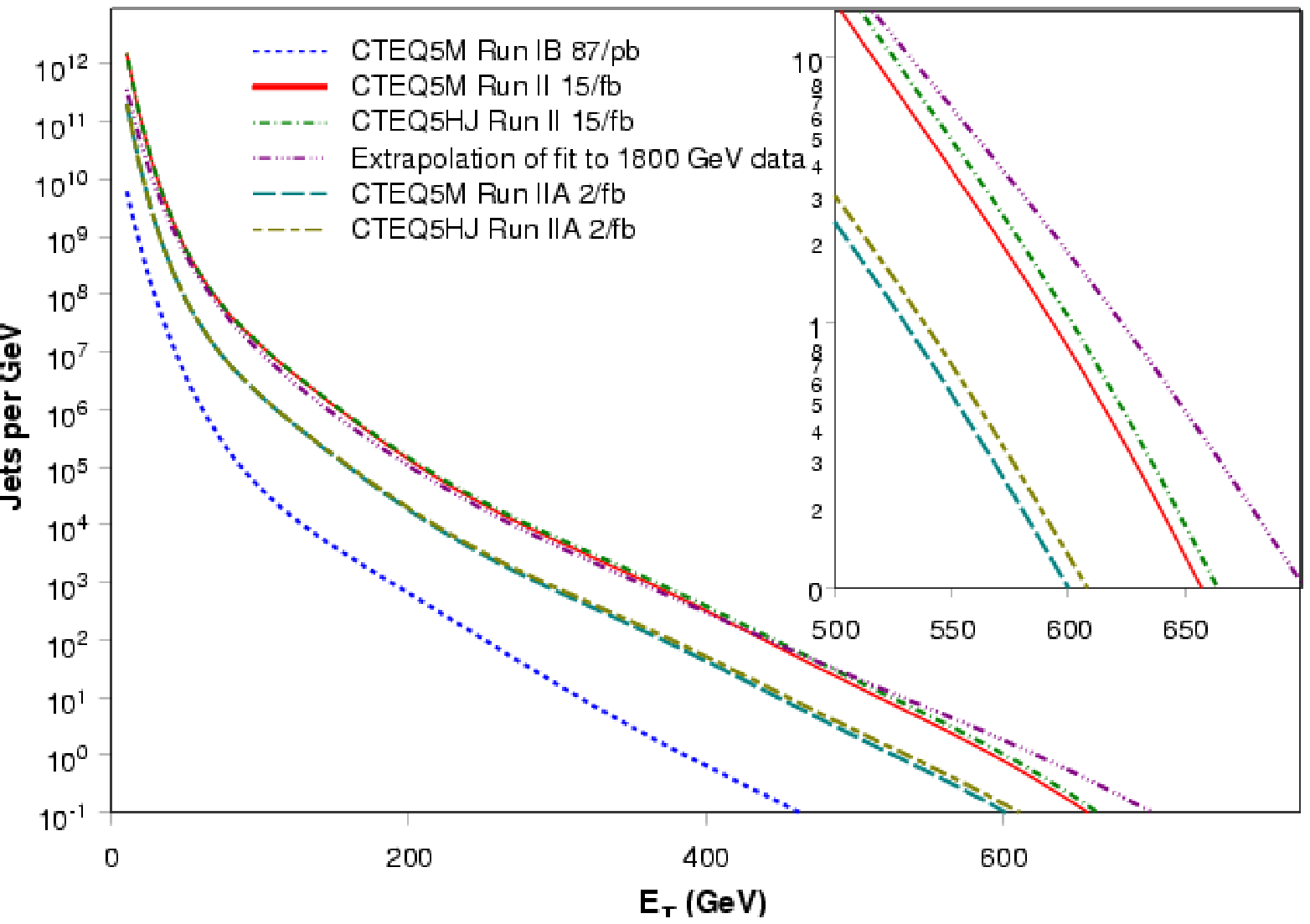,height=6.8cm,width=7cm}}}
\caption{(left)The dijet differential  cross section for CDF from Run
1B;
(right) The inclusive jet cross section (number of events) in the
central rapidity region for CDF.  Predictions for Run 1b, Run 2a, Run
2b using EKS NLO program and the CTEQ5M and CTEQ5HJ PDFs, as well as
an extrapolation of the physics curves measured by CDF in Run 1b.}
\label{dijet}
\end{figure}

The rapidity dependence of the dijet differential  cross section
provides information about new regions of $x$ and $Q^2$.  Assuming a
2$\rightarrow$2 hard scattering, the event kinematic variables
$(x,Q^2)$ are related to the jet's transverse energy $E_T$ and
pseudorapidity $\eta$ as
\[x_{12}=\Sigma_{i}\frac{E_{T_i}}{\sqrt{s}}\exp^{\pm\eta_i};
\;\;\;\;\;Q^2=2{E_T}^2\cosh^2\eta^{\ast}(1-\tanh\eta^{\ast}),\]
where the sum is over all jets in the event.  The $x_1,x_2$ are parton
momentum fractions and $Q^2$ is the two body four momentum transfer of
the interaction.  CDF measured the dijet differential 
cross section~\cite{dijet} 
where one jet is restricted to the central 
$(0.1\le\mid\eta\mid\le 0.7)$ region and the other jet 
is constrained to  different rapidity
regions.  Fig.~\ref{dijet}~(left) shows the CDF data compared to the
theoretical predictions.  An excess similar to that of the
inclusive jet cross section is observed. 
%--------------------------------------------------------------------------
%----------------------------------------------

\section{Run 2 Prospects}

One of the goals of the Run 2 is to obtain higher precision for QCD
measurements. 
A data sample of 2$fb^{-1}$ will enable the jet cross section to be
probed for higher $E_T$ and $x$ values, than were possible at Run 1. 
In addition to the high statistics the 
increase in the center-of-mass energy from 1800 GeV to 1960 GeV 
has a large effect on the high $E_T$ jet rate. For the full Run 2A
sample the numbers of jets above 400 GeV are expected to increase from
11 to $\sim 500$, with energy reach extending to 550 GeV, 
see Fig.~\ref{dijet}~(right). For jet identification 
CDF will use both standardized  jet cone and $K_T$ algorithms.  

The  Run 2 calorimeter upgrade  provides calorimetry
as precise in the forward region as in the central one.
The inclusive jet cross section will be measured out to rapidity
values of 3 in Run 2, and dijet differential  measurement can be
enhanced by allowing  both jets to be non-central. 
In Run 2 the measurement of the strong coupling constant will be possible
from a $Q^2$ of 100 GeV$^2$ to 300000 GeV$^2$.
With more luminosity delivered by Tevatron every day, 
we are looking forward to new  QCD precision measurements at CDF.

\vglue.6cm
{\bf Acknowledgments}
\vglue.3cm
It is a great pleasure  to thank Prof. G. Zoupanos and organizers 
for opportunity to give this talk  
and for their very warm hospitality in Corfu.


\begin{thebibliography}{99}

\bibitem{CDF-detector}F. Abe {\it et~al}., \Journal{\NIMA}{271}{387}{1988}.
\bibitem{CDF-Run1b}T. Affolder {\it et~al}., \Journal{\PRD}{64}{032001}{2001}.
\bibitem{dijet-mass}T. Affolder {\it et~al}., \Journal{\PRD}{61}{091101}{2000}. 
\bibitem{dijet}T. Affolder {\it et~al}., \Journal{\PRD}{64}{012001}{2001}.
\bibitem{sumet}F. Abe {\it et~al}., \Journal{\PRL}{80}{3461}{1998}.
\bibitem{CTEQ4M}H.~L. Lai {\it et~al}., \Journal{\PRD}{55}{1280}{1997}.
\bibitem{Walter-alphas}W. Giele {\it et~al}., 
%E.~W.~N. Glover, J. Yu, 
\Journal{\PRD}{53}{120}{1996}.
\bibitem{Ellis-rsep}S.~D. Ellis {\it et~al}., 
%Z. Kunszt, and D.~E. Soper, 
\Journal{\PRL}{69}{3615}{1992}.
\bibitem{JETRAD}W. Giele {\it et~al}., 
%E.~W.~N. Glover, D.~A. Kosower, 
\Journal{\PRL}{73}{2019}{1994}.
\bibitem{alphas-prl}T. Affolder {\it et~al}., \Journal{\PRL}{88}{042001}{2002}.
\bibitem{MRST}A.~D. Martin {\it et~al}., \Journal{\EPJ}{4}{463}{1998}.
\bibitem{MRSA}A.~D Martin {\it et~al}., 
%R. Roberts, and W.~J. Stirling, 
\Journal{\PLB}{356}{89}{1995}.
\bibitem{MRSR}A.~D. Martin {\it et~al}., 
%R. Roberts, and W.~J. Stirling, 
\Journal{\PLB}{387}{419}{1996}.
\bibitem{world}D.~E. Groom {\it et.al}., Particle Data Group, 
\Journal{\EPJ}{15}{1}{2000}.

\end{thebibliography}
\end{document}